\documentclass{iaus}
\usepackage{graphicx}

\title[Near-surface stellar magneto-convection] %% give here short title %%
{Near-surface stellar magneto-convection: simulations for the Sun
and a metal-poor solar analog}

\author[M. Steffen, H.-G. Ludwig \& O. Steiner]   %% give here short author list %%
{Matthias Steffen$^1$,
 H.-G. Ludwig$^2$,  \and O. Steiner$^3$}

\affiliation{$^1$Astrophysikalisches Institut Potsdam, An der Sternwarte 16,
D-14482 Potsdam, Germany,\break email: msteffen@aip.de\\[\affilskip]
$^2$GEPI, Observatoire de Paris, CNRS, Universit\'e Paris Diderot,
92195 Meudon Cedex, France\\$^3$Kiepenheuer-Institut f\"ur
Sonnenphysik, Sch\"oneckstrasse 6, D-79104 Freiburg, Germany}

\pubyear{2009}
\volume{259}  %% insert here IAU Symposium No.
\pagerange{119--126}
\date{Dec.30,2008 and in revised form ??}
 \jname{Cosmic Magnetic Fields: From Planets,
to Stars and Galaxies} \editors{K.G. Strassmeier, A.G. Kosovichev \&
J.E. Beckman, eds.}
\begin{document}

\maketitle

\begin{abstract}
We present 2D local box simulations of near-surface radiative
magneto-convection with prescribed magnetic flux, carried out with
the MHD version of the CO$^5$BOLD code for the Sun and a solar-like
star with a metal-poor chemical composition (metal abundances
reduced by a factor 100, [M/H]=--2). The resulting
magneto-hydrodynamical models can be used to study the influence of
the metallicity on the properties of magnetized stellar atmospheres.
A preliminary analysis indicates that the horizontal magnetic field
component tends to be significantly stronger in the optically thin
layers of metal-poor stellar atmospheres. 
%, Sun: magnetic fields, stars: magnetic fields}
\end{abstract}

\firstsection % if your document starts with a section,
\section{Introduction}
\label{sec:introduction}
Up to now, numerical simulations of near-surface magneto-convection
are essentially restricted to the Sun
(see e.g.\ \cite[Carlsson et al.\ 2004]{carlssonetal04};
\cite[V\"ogler et al.\ 2005]{voegleretal05};
\cite[Stein \& Nordlund 2006]{sn06};
\cite[Sch\"ussler \& V\"ogler 2008]{sv08};
\cite[Steiner et al.\ 2008]{steineretal08}).
The idea of the present work is to study the influence of
metallicity on the structure and distribution  of magnetic flux
concentrations in stellar atmospheres. For this purpose,
realistic local box simulations of near surface magneto-convection
with prescribed magnetic flux have been carried out for the Sun and a
metal-poor solar analog, allowing a differential comparison of the
thermal, dynamical, and magnetic properties of the photospheric
layers.\\[-8mm]
\section{2D Numerical simulations}
\label{sec:simulations} 
\begin{figure}
 \mbox{\includegraphics[bb=14 56 570 370,width=0.5\textwidth]
{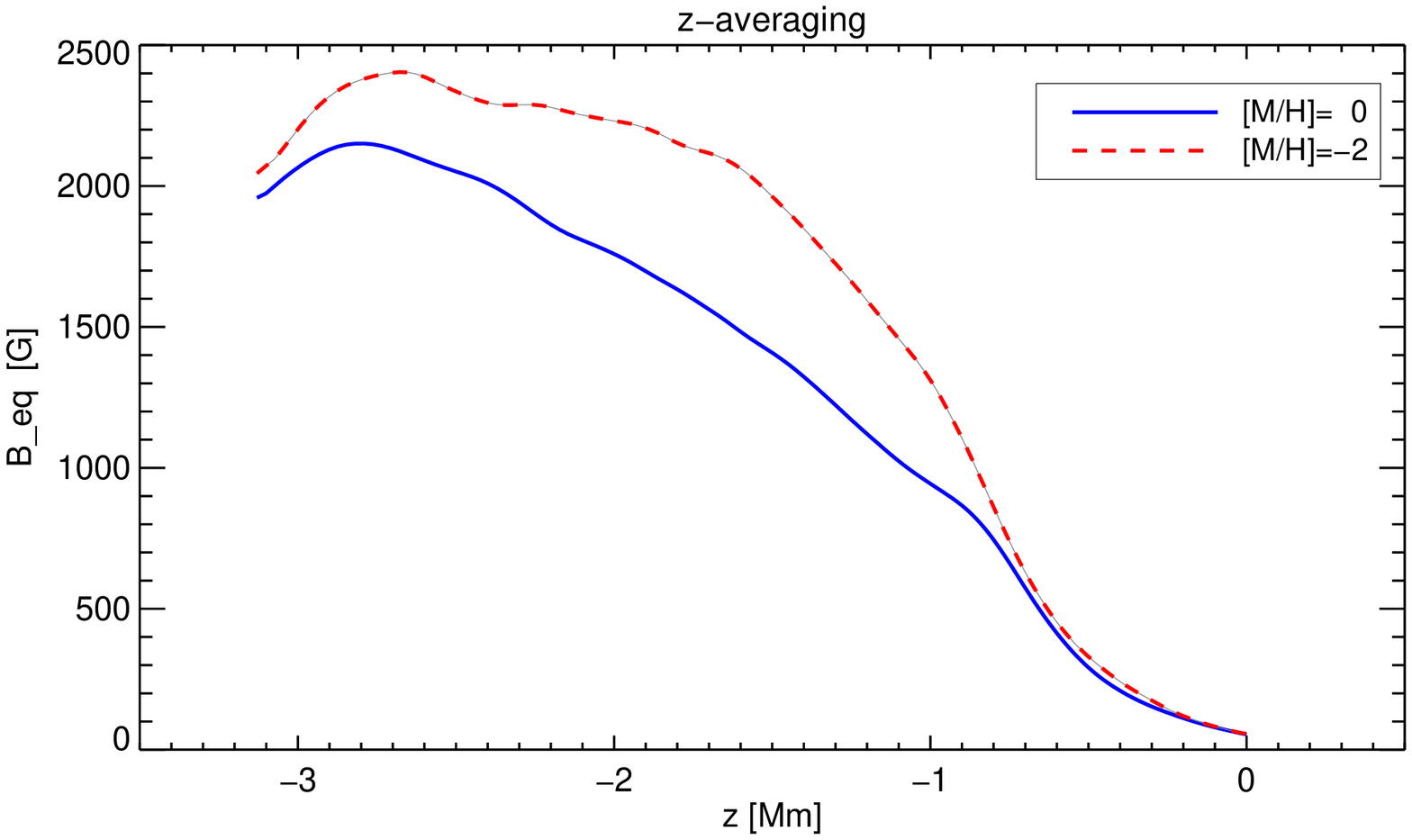}}
 \mbox{\includegraphics[bb=14 56 570 370,width=0.5\textwidth]
{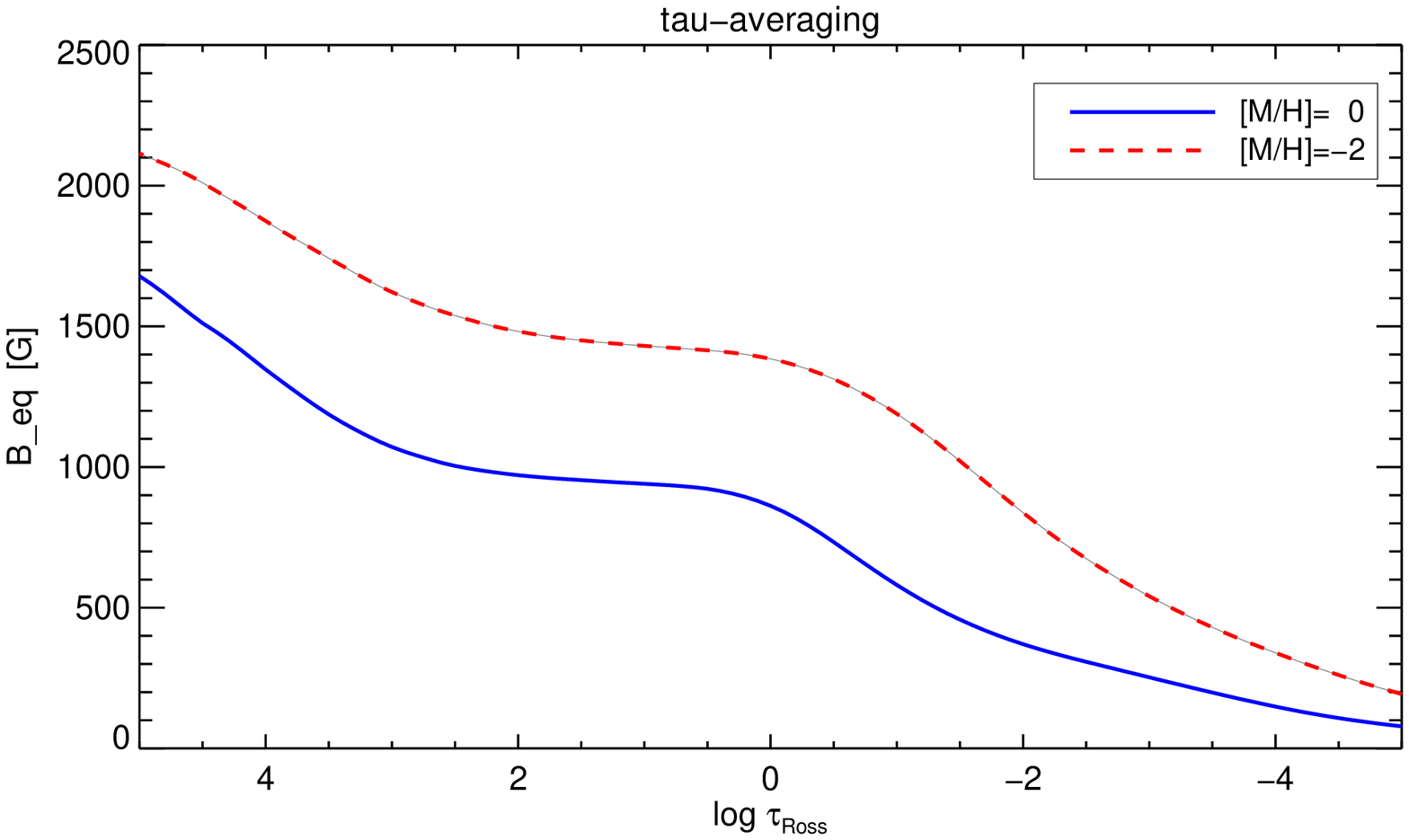}}
 \caption{Equipartition field strength, $B_{\rm eq}$ (see text), on the
   geometrical height scale (left) and on the optical depth scale (right),
   computed from the two non-magnetic simulations.}
\label{fig:1}
\end{figure}

We compare two sets of simulations, both
with $T_{\rm eff}$\,$=$\,$5690$~K, and $\log g$\,$=$\,$4.44$, but
with different chemical composition ([M/H]=0 and [M/H]=$-2$). All
models are computed on the same 2D Cartesian grid (400 $\times$ 165
cells, $\Delta x$\,$=$\,$28$~km, $12$~km $<$ $\Delta z$ $<$ $28$~km,
periodic side boundaries and open/transmitting boundaries at
bottom/top). Gray radiative transfer with realistic opacities for
the respective chemical composition are used together with an
appropriate equation of state that accounts for partial ionization
of H and He. Note that the main effect of the reduced metallicity is
a general decrease of the opacity; as a consequence, the gas
pressure at the visible surface ($\tau$\,$=$\,$1$) is about $3$
times higher in the metal-poor model. Figure \ref{fig:1} shows the
equipartition magnetic field strength $B_{\rm eq} \equiv
\sqrt{4\pi\langle\rho\vec{u}^2\rangle}$ (where
$\langle\rho\vec{u}^2\rangle/2$ is the horizontally averaged kinetic
energy density) on the geometrical and optical depth scale, as
computed from the two non-magnetic simulations.

A homogeneous unipolar vertical magnetic field of four different strengths
($B_{\rm z,0}$\,$=$\,$10$ (quiet) $20$, $40$, and $80$ (network) Gauss,
respectively, was superimposed on the non-magnetic models, defining the
initial conditions for a series of ideal simulations that are used to study
near-surface magneto convection at different metallicity as a function of
the prescribed mean magnetic flux. At top and bottom, the magnetic field
is forced to be vertical ($B_{\rm x}$\,$=$\,$B_{\rm y}$\,$=$\,$0$);
at each height, the mean vertical magnetic flux thus retains its initial value
$\langle B_{\rm z}\rangle$\,$=$\,$B_{\rm z,0}$ throughout these
\emph{ideal} ($\eta$\,$=$\,$0$) MHD simulations.

\begin{figure}
 \mbox{\includegraphics[bb=0 0 496.949 281,width=0.5\textwidth]
 {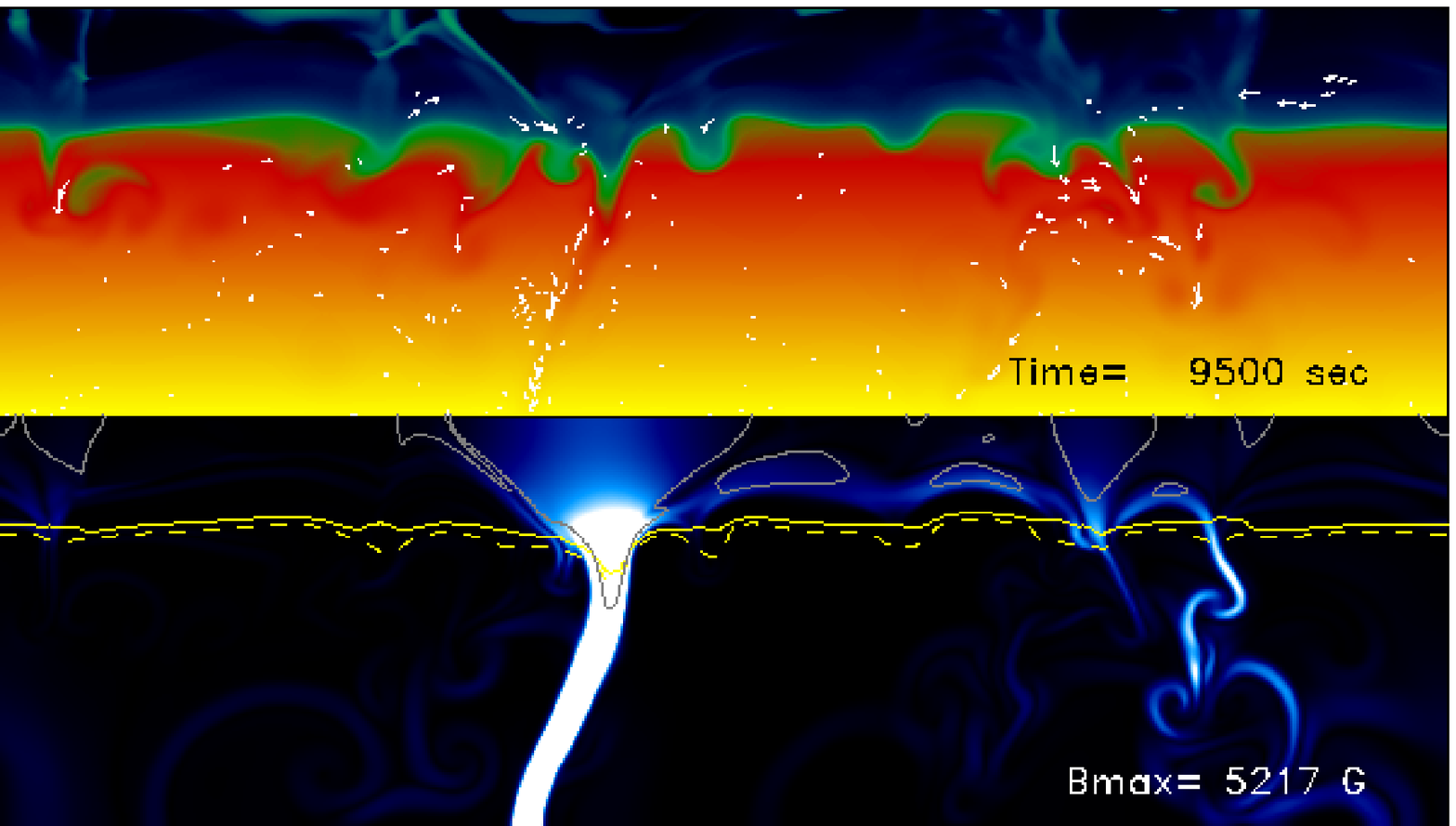}}
 \mbox{\includegraphics[bb=0 0 496.949 281,width=0.5\textwidth]
 {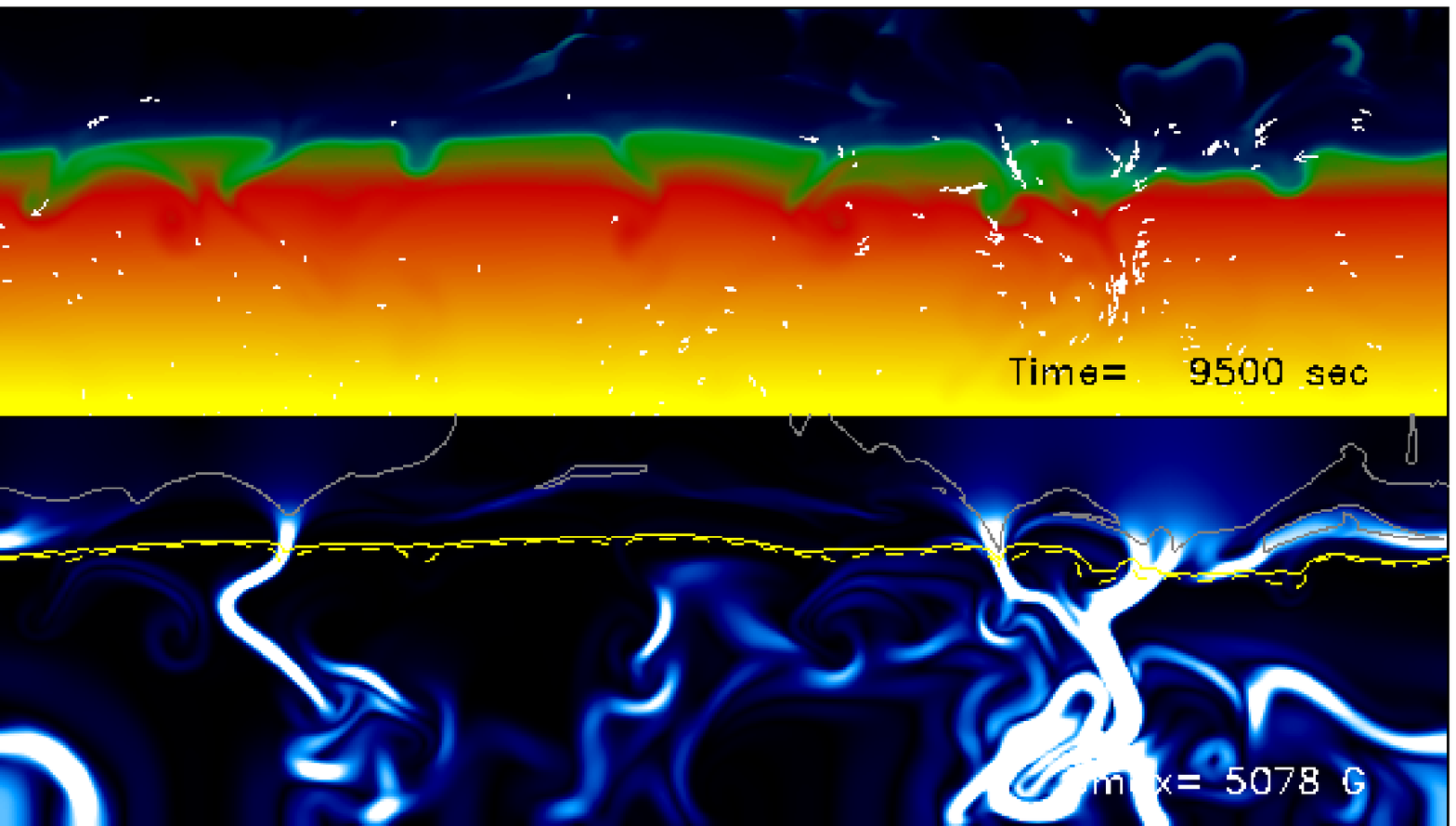}}
 \caption{
 Snapshots from 2D\,MHD simulations with solar (left) and metal-poor (right)
 composition, taken 9\,500\,s after inserting the initial magnetic field
 ($B_{\rm z,0}$\,$=$\,$80$~G). Upper panels: temperature structure,
 lower panels: distribution of magnetic field density $|B|$; the dashed
 (yellow) line delineates the $\tau$\,$=$\,$1$ contour, gray lines are
 $\beta$\,$=$\,$1$ contours. $|B|_{\rm max} \approx 1.5$~kG at
 $\tau$\,$=$\,$1$.}
\label{fig:2}
\end{figure}

An illustration of the resulting magnetic filed configurations is given
in Fig.\,\ref{fig:2}.
A first inspection of the simulation data suggests that, for given
$B_{\rm z,0}$, the rms vertical field $B_{\rm z,rms}$ evaluated at
fixed optical depth is very similar for the two metallicities, even though
$B_{\rm eq}$ is significantly different (Fig.\,\ref{fig:1}). However,
there is a tendency for stronger horizontal fields in the photosphere of
the metal-poor simulations. \\[-8mm]


\section{Conclusions}
These preliminary results need to be confirmed by more extensive 2D
and 3D simulations with non-gray radiative transfer and alternative
MHD boundary conditions. In the long run, further questions can be
addressed by computing synthetic (Stokes) spectra based on the MHD
model atmospheres, e.g.\ concerning the influence of small-scale
surface magnetic fields on the micro-variability of the stellar
radiation and/or on the accuracy of spectroscopic abundance
determinations. 

\end{document}